\documentclass[12pt]{iopart}

\usepackage{graphicx}
\usepackage{amssymb}
\usepackage{setspace}
\usepackage{color,soul}
\usepackage{tabularx}
\usepackage{array}

\begin{document}

\title[]{A method to optimize patient specific HU-SP calibration curve from proton range measurements.}

\author{Jaroslav Albert$^1$, Rudi Labarbe$^2$, Edmond Sterpin$^{3,4,5}$}
\vspace{10pt}
\address{$^1$Universit\'e Catholique de Louvain, ELEN
Place du Levant 2/L5.04.04
1348 Louvain-la-Neuve ,Louvain-la-Neuve, Belgium}
\ead{jaroslav.albert@uclouvain.be}
\vspace{10pt}
\address{$^2$IBA Particle Therapy, Chemin du Cyclotron 3, B-1348 Louvain-la-Neuve, Belgium}
\ead{rudi.labarbe@iba-group.com}
\vspace{10pt}
\address{$^3$Universit\'e Catholique de Louvain, Molecular Imaging Radiotherapy and Oncology, Brussels, Belgium
}
\address{$^4$
Universit\'e Catholique de Louvain, Institut de Recherche Exp\'erimentale et Clinique, Brussels, Belgium
}
\address{$^5$
KU Leuven, Department of Oncology, Laboratory of Experimental Radiotherapy, Leuven, Belgium
}
\ead{edmond.sterpin@uclouvain.be}
\vspace{10pt}
\begin{indented}
\item[]

\end{indented}

\begin{abstract}
Proton radiotherapy promises accurate dose delivery to a tumor and minimal dose deposition to all
other tissues. However, in practice the planned dose distribution
may not conform to the actual one due to noisy data and different types of errors. 
One such error comes in a form of potentially inaccurate conversion of the Hounsfield units (HU) to
stopping powers (SP) of protons.
We propose a method of improving the CC based on a planning CT and
proton range measurements acquired during treatment.
The range data were simulated using a virtual CC and a planning CT. The range data were given two types of noise:
range shift due to patient setup errors; and range noise due to measurement imprecision, including a misalignment of
the range measuring device.
The method consists of two parts.
The first part involves a Taylor expansion
of the water equivalent path length (WEPL) map in terms of the range shift caused by the difference between the planning and the virtual CC.
The range shift is then solved for explicitly, leading to a polynomial function of the difference between the two CCs.
The second part consists in minimizing a score function relating the range due to the virtual CC and the range due to the optimized CC.
Tested on ten different CCs, our results show that, with range data collected over a few fractions (less than 10), the optimized
CC leads to an overall reduction of the range difference. More precisely, on average, the uncertainty of the CC was reduced from 2.67\% to 1.62\%, while
the average reduction of the WEPL bias was reduced from 2.14\% to
0.74\%.
The advantage of our method over others is 1) its speed, and 2) the fact that the range data it necessitates are acquired
during the treatment itself, and as such it does not burden the patient with additional dose.
   
\end{abstract}

%
\vspace{2pc}
\maketitle
%
%

\section{Introduction}

Proton radiotherapy has the potential of delivering precise radiation dose to tumors while sparing healthy tissue more than
conventional radiotherapy, especially in the
distal area behind the tumor. Hindering this potential is protons' sensitivity to errors, which may include 
changes in the anatomy due to inter/intra-fraction motion \cite{Langen, Young}, 
patient set up errors \cite{Letourneau}, 
but also
imperfect translation of patient image data into dosimetric quantities. The latter includes: noise in the Hounsfield Units (HUs) of the CT scan \cite{Fleischmann, Chvetsov}, the 
conversion of the HUs to stopping power (SP) \cite{Paganetti} and artifacts in the CT. 
Although these errors may be small, their collective contribution can lead to a significant over or under-range
inside a patient.

The HUs can be converted into SPs through 
various methodologies, either via single \cite{Schaffner, Jiang, Yang, Schneider, Moyers} or dual/multi energy 
\cite{Farace, Saito, Grantham} CT scanners. Both of these methods rely on a stoichiometric approximation \cite{Schaffner, Schneider, Bourque}
relating the HUs to the relative stopping power (RSP), which is defined as the stopping power (SP) of a medium (e. g. tissue) divided
by the SP of water. 
However, uncertainties in the RSP remain due to inaccurate tissue segmentation 
(for instance, because of the noise) and inherent uncertainties 
linked to the mean excitation energies (I value) needed to 
compute stopping powers \cite{Doolan1}.

A common way to relate the RSP to HUs is via a calibration curve (CC). 
In recent years, a patient-specific approach to estimating the CC has been in development, one in which range data from an individual patient are collected
via proton radiography (PR). In this approach, a patient is irradiated with protons of sufficient energy to traverse the patient and their Bragg profiles are
detected by either a Multi-layer ionization chamber (MLIC), an integrating detector or a proton-by-proton counting device.
From these profiles, it is possible to optimize the stoichiometric CC 
in order to be more suitable for the patient \cite{Doolan}. 
Alternative to the PR is a another invention for measuring the actual (clinical) range inside a patient \cite{Xie, Priegnitz, Smeets, Xing, Nenoff, Janssens}.
Known as Prompt gamma camera (PGC), this device detects gamma radiation coming from nuclear reactions between the protons and the atomic nuclei 
whereby allowing one to reconstruct the Bragg profile from which the range
can be determined. There exist other devices/methods, such as PET scan and ionoacoustic measurements \cite{Kellnberger,Lehrack}, 
that are able to measure, among other quantities, the proton range. In this paper, we
do not focus on any specific device; we merely take as granted that range data can be collected during a treatment session.

What we are proposing is a method of optimizing the patient-specific CC that incorporates 
range measurements collected during a proton radiotherapy session.
The method involves two main steps.
The first step is dependent on the planning CT only, 
and can be carried out before the range measurements.
Consequently, the second step, which leads to
the improvement of the CC and is performed as soon as the range data are collected, can be performed in
a matter of seconds.
In order to alleviate the problem of multiple minima, engendered by noise/imperfect data, we
place constraints on the way in which the CC can be adjusted.
First, we partition the CC into sections corresponding to the highest frequency HU.
Secondly, we restrict the deviations from the stoichiometric CC to be within 5$\%$ and require that the optimized CC increase monotonically.

The novelty of this method is two fold. The first is its speed. While a more accurate approach would involve a Monte Carlo
simulation of the pencil beams (PBs), the time frame for such an approach would not be practical, as the input of the simulation
is not one or two parameters, but an entire function, i. e. the CC. The second is its adaptability to any range measuring device/method.
The source of these two novelties is what lies at the core of our method: an analytical expression relating the planning CT and the range measurements
to the CC.

\section{Materials and Methods}

Our goal is to optimize the calibration curve (CC) using only the planning CT and range measurement data.
The CC used at the planning stage will be referred to as CC$^{pl}$, while the optimized CC will be labeled
as CC$^{opt}$. For testing purposes, we also define a virtual CC, CC$^{v}$, which represents an ideal CC, one that, if
used instead of CC$^{pl}$, would produce a range map that is as close to the real range map as possible.
 In the context of this definition, the goal here is
to modify CC$^{pl}$ so as to make CC$^{opt}$ as close to CC$^{v}$ as possible.

We begin by computing the water-equivalent path length (WEPL) map from CC$^{pl}$. The WEPL is defined as the distance a pencil beam would travel if the medium of 
interest, e. g. tissue, was replaced by water. For a proton moving along the $z$-axis, the WEPL map $W(x,y,z)$ can be computed using 
the continuos slowing down approximation (CSDA) sheme
\cite{Newhauser},
which yields the formula
\begin{equation}\label{WEPL}
W(x,y,z)=\int_0^zCC(H(x,y,z'))dz',
\end{equation}
where $x, y$ are the coordinates in the plane perpendicular to the proton's path and
$CC(H(x,y,z))$ is the calibration curve as a function of $H$. 
Note that $CC$ does not depend on the beam's energy, for the energy range applicable to proton therapy \cite{Schneider}.
Once we have the WEPL map, we can compute the range $R(x,y)$ of a pencil beam of energy $E$ passing through any given pixel $(x,y)$ by solving
for $z$ in Eq. (\ref{WEPL}).
If $CC=$CC$^{v}$, $R(x,y)$ will correspond to the virtual range. If, however, $CC=$CC$^{pl}$, and CC$^{pl}\neq$CC$^{v}$, 
there will be a discrepancy, $\Delta R(x,y)$, between the
virtual range $R^{v}(x,y)$ and the one computed using CC$^{pl}$, $R^{pl}(x,y)$: $\Delta R(x,y)\equiv R^{v}(x,y)-R^{pl}(x,y)$.

We proceed by noting that whatever CC$^{v}$ may be, it will not be different from CC$^{pl}$ by more than 
a few percent (max $\sim 5\%$). This means that the range computed using CC$^{pl}$, $R^{pl}$, and the one arising from CC$^{v}$, $R^{v}$, 
will also differ by only a few percent;
that is, $(R^{v}-R^{pl})/R^{v}<<1$. This allows us to expand the WEPL map computed from CC$^{v}$, which will be
denoted as $W^{v}$, in $\Delta R$:
\begin{eqnarray}\label{WvirtExpanded}
W^{v}(x,y,R^{v})&=&W^{v}(x,y,R^{pl})
+\frac{d}{dz}W^{v}(x,y,z)\bigg|_{z=R^{pl}}\Delta R\nonumber\\
&+&\frac{1}{2}\frac{d^2}{dz^2}W^{v}(x,y,z)\bigg|_{z=R^{pl}}\Delta R^2+...\,.
\end{eqnarray}
Let us examine the magnitudes of the second and third term. The former is a first derivative of $W^{v}$ with respect to
$z$, which is merely CC$^{v}$ (see Eq. (\ref{WEPL})), multiplied by $\Delta R$. The CC is of order 1, so this term will be or order $\Delta R$.
The latter is a derivative of CC$^{v}$, multiplied by $(\Delta R)^2$. On average, the derivative of a CC is approximately
the maximum value of the CC divided by the total range of HU: $CC(3000)/(3000-(-1000))\sim 10^{-3}$. Hence, the
third term is of order $10^{-3}\Delta R^2$, which is smaller than the second term by a factor of $\sim10^{-3}\Delta R$. Even for $\Delta R$
as large as 10 mm, this term can be safely neglected.
Solving for $\Delta R$, we obtain
\begin{equation}\label{DeltaR}
\Delta R=\frac{W^{v}(x,y,R^{v})-W^{v}(x,y,z)}{dW^{v}(x,y,z)/dz}\bigg|_{z=R^{pl}}.
\end{equation}
Since CC$^{v}$ will differ from CC$^{pl}$ by only a few percent, so will $W^{v}$ from $W^{pl}$. Thus, we can express $W^{v}$ as
\begin{equation}
W^{v}(x,y,z)=W^{pl}(x,y,z)+\phi(x,y,z),
\end{equation}
where $\phi(x,y,z)=W^{v}(x,y,z)-W^{pl}(x,y,z)\ll W^{v}(x,y,z)$,
and then expand
the right hand side of Eq. (\ref{DeltaR}) in powers of $d\phi/dz=\phi'$:
\begin{eqnarray}\label{phi}
\Delta R&=&\frac{W^{v}(x,y,R^{v})-W^{v}(x,y,z)}{dW^{v}(x,y,z)/dz}\bigg|_{z=R^{pl}}\nonumber\\
&=&\frac{W^{v}(x,y,R^{v})-W^{pl}(x,y,R^{pl})-\phi(x,y,R^{pl})}{CC^{pl}+\phi'(x,y,R^{pl})}\nonumber\\
&=&\frac{-\phi(x,y,R^{pl})}{CC^{pl}(x,y,R^{pl})}+\frac{\phi(x,y,R^{pl})\phi'(x,y,R^{pl})}{(CC^{pl}(x,y,R^{pl}))^2}+...,
\end{eqnarray}
where we set $dW^{pl}/dz=$CC$^{pl}$. The function $\phi$ is a correction to $W^{pl}$, while $\phi'$ is a correction to CC$^{pl}$.
The term $W^{v}(x,y,R^{v})-W^{pl}(x,y,R^{pl})$ vanishes
thanks to the relation $W^{v}(x,y,R^{v})=W^{pl}(x,y,R^{pl})$, or, more explicitly:
\begin{equation}
\int_0^{R^{v}}CC^{v}(x,y,z')dz'=\int_0^{R^{p}}CC^{pl}(x,y,z')dz'
\end{equation}
One way to construct $\phi'$ is in terms of some bases functions $\xi_j(x,y,z)$ such that
\begin{equation}\label{phiForm}
\phi'(x,y,z)=\sum_{j=0}^Mu_j\xi_j(x,y,z),
\end{equation}
with $u_1, u_2,..., u_M$ being some parameters.
This allows Eq. (\ref{phi}) to be written in the form
\begin{equation}\label{phiConcise}
\Delta R=-\sum_{j=0}^Mu_j\alpha_j(x,y)+
\sum_{j=0}^M\sum_{k=0}^Mu_ju_k\beta_{jk}(x,y)+...,
\end{equation}
where
\begin{eqnarray}\label{alphabeta}
&&\alpha_j(x,y)=\int_0^{R^{pl}}\frac{\xi_j(x,y,z')}{CC^{pl}(x,y,R^{pl})}dz'\nonumber\\
&&\beta_{jk}(x,y)=\int_0^{R^{pl}}\frac{\xi_j(x,y,z')\xi_k(x,y,R^{pl})}{(CC^{pl}(x,y,R^{pl}))^2}dz'
\end{eqnarray}
Note that the parameters $u_1,...,u_M$ in Eq. (\ref{phiConcise}) are outside of the brackets, which means that the brackets can be computed
in advance, before optimizing the parameter set  $u_1,...,u_M$. 

Eq. (\ref{phi}) assumes that all the protons of a pencil beam pass through the pixel $(x,y)$. In reality this is far from true; the beam has
a lateral Gaussian spread with standard deviation of approximately 3 mm, 
which means that protons as far away as 7 mm from $(x,y)$
contribute to the shape of the Bragg curve. In two recent studies \cite{Farace2, Deffet}, it has been
demonstrated that a simple convolution over a Gaussian with a $\sigma$ equal to the width of the
pencil beam provides an excellent approximation of the dose profile, and hence the average range.
One of the upshots of these studies is that multiple Coulomb scattering can be neglected.
Thus, in order to take into account the width of the pencil beams,
we must perform a convolution of both sides of Eq. (\ref{phiConcise}), $(\Delta R*G_{\sigma})(x,y)$
where $G_{\sigma}$ is a Gaussian centered at (x,y) with standard deviation $\sigma$, and the operator $*$ means
\begin{equation}\label{convo}
(f*G)(x,y)=\frac{1}{2\pi\sigma^2}\int_{-\infty}^{\infty}f(x',y')e^{-[(x'-x)^2+(y'-y)^2)]/2\sigma^2}dx'dy'.
\end{equation}
Thus, we obtain
\begin{equation}\label{phiconvol}
(\Delta R*G_{\sigma})(x,y)=
-\sum_{j=0}^Mu_j\left(\alpha_j*G_{\sigma}\right)(x,y)
+\sum_{j=0}^M\sum_{k=0}^Mu_ju_k\left(\beta_{jk}*G_{\sigma}\right)(x,y).
\end{equation}
where $(\Delta R*G_{\sigma})=(R^{v}*G_{\sigma})-(R^{pl}*G_{\sigma})$. The quantity $(R^{v}*G_{\sigma})$ represents 
the measured range of a pencil beam centered at $(x,y)$ and will be denoted as $R^m$.
In order to compute $(R^{pl}*G_{\sigma})$, we must solve Eq. (\ref{WEPL}) with $CC^{pl}$ for all pixels $(x, y)$,
interpolate the solution in the x-y plane, and then perform the convolution according to Eq. (\ref{convo}).

\subsection{Application to proton therapy: head and neck}

Let us now apply the techniques developed in the previous section to a patient. Figures 1 a) and b) show the CT and the dose from one beam (out of three) of a head and neck patient. 
The CT and the treatment plan were obtained from Cliniques universitaires Saint-Luc, MIRO.
Since we do not know in advance the dependence of our range measuring device on the intensity of the PBs, we
must tets our methods only on PBs of the same intensity (or weight). In order to maximize the number of data points, we chose a group of
PBs with weights between 0.015 and 0.025 shown as the black columns in Figure 1 c). Also shown are PBs from
the chosen group for layers 8, 10 and 16.

Next, we must construct a CC$^v$ that will serve as an ideal CC. We do this by introducing deviations to
CC$^{pl}$, which was obtained from Cliniques universitaires Saint-Luc, MIRO. 
To ensure that the deviations are smooth and correlated, we first add noise to the derivative of
CC$^{pl}$ sampled from a simple birth-death process, and then integrate the noisy $d$CC$^{pl}/dHU$
up to $HU$. Figure 2 a) shows ten CCs thus generated. 
As shown in Figure 2 b), by far the most frequent HUs in a volume of tissue surrounding 
the PTV are those near $-1000$ and near zero, e. g. $-120$ to $200$. Since we are not interested in correcting
the CC for air ($HU\sim -1000$), we will focus only on the region  $[-120,200]$.

\begin{figure}
\vspace*{3cm}
\includegraphics[trim=0 0 0 5.0cm, height=0.6\textheight]{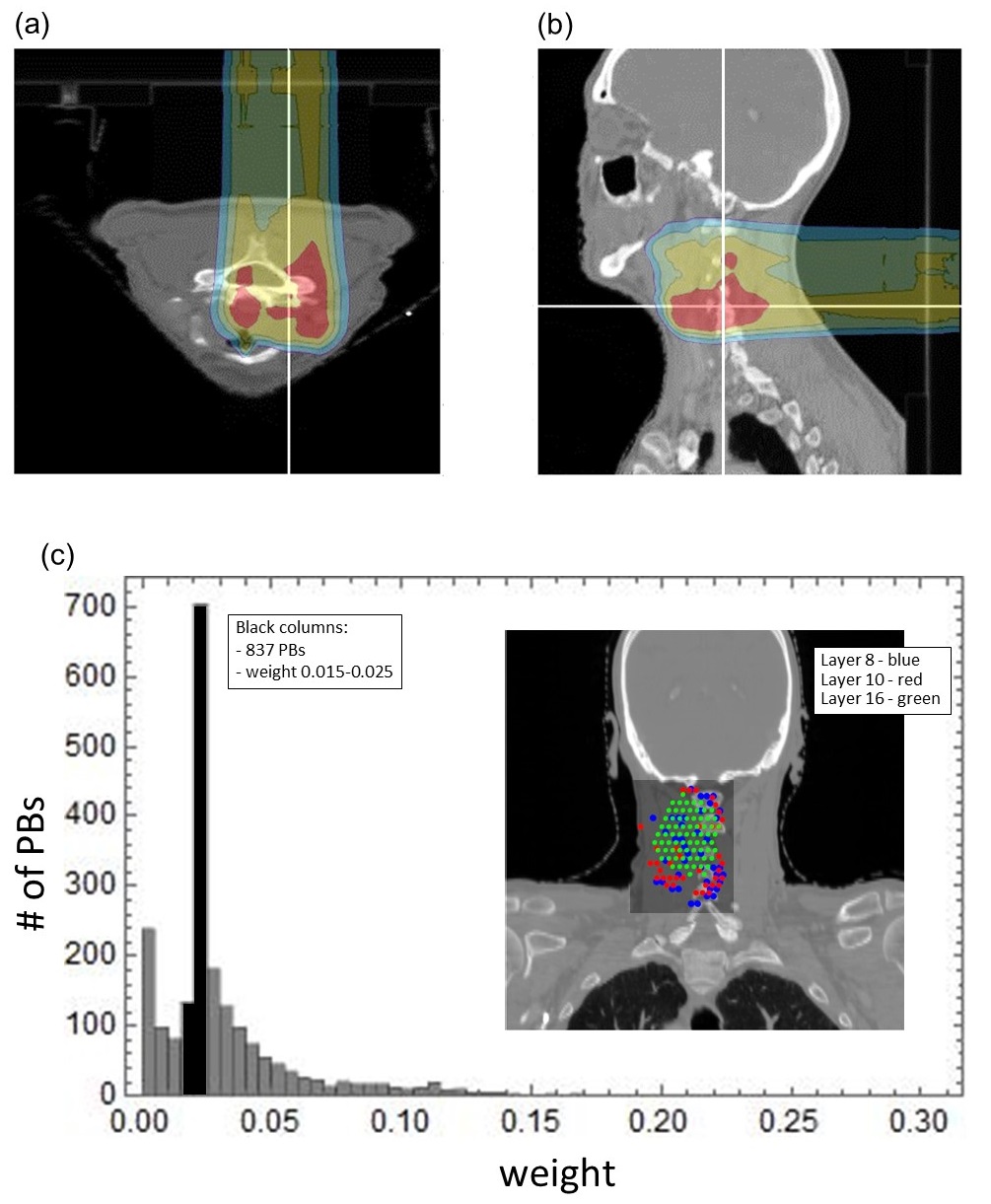}
\centering
\caption{a) and b) A CT of a patient and the dose simulated by Monte Carlo. c) A histogram showing the
distribution of PB weights. The black columns indicate the weights that were chosen for optimization.
The CT on the right shows the PBs with the weight range of $0.015 - 0.025$ 
for three energy layers: blue dots = 113.72 MeV; red dots = 107.77 MeV; and green dots = 92.02 MeV.}
\end{figure}

\begin{figure}
\includegraphics[trim=0 0 0 0.5cm, height=0.41\textheight]{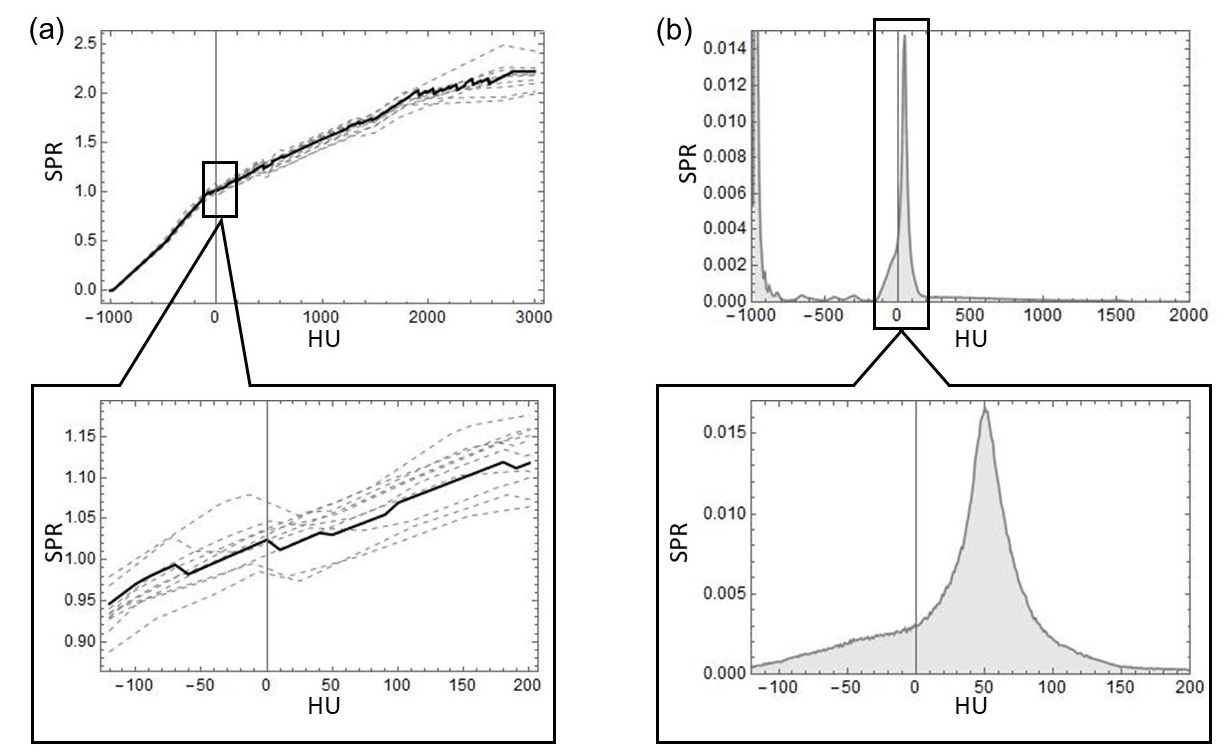}
\caption{a) Ten virtual CCs (dashed gray lines) and the planning CC (solid black line).
b) Distribution of HUs. The vertical axis was normalized such that the area between -120 and 200 is one.}
\end{figure}

Let us now chose the structure of $\phi'$. The simplest choice might be a piece-wise linear function:
\begin{equation}\label{phiLinear}
\phi'(x,y,z)=\sum_{j=0}^{M-1}(m_jH(x,y,z)+b_j)\Omega_j(H(x,y,z)),
\end{equation}
where 
\begin{equation}
m_j=\frac{u_{j+1}-u_j}{h_{j+1}-h_j},
\end{equation}
is the slope of the line in the section $h_j\leq H<h_{j+1}$ and
\begin{equation}
b_j=\frac{h_{j+1}u_j-h_ju_{j+1}}{h_{j+1}-h_j}.
\end{equation}
is the y-intercept. The set of coefficients $h_1,...,h_M$ represent
$HU$s on the interval $[-120,200]$. The function $\Omega_j$ is defined as
\begin{equation}
\Omega_j(x)=\Bigg\{
\normalsize\begin{array}{c}
1\,\,\,\,\,\,\,\,\,\,\textrm{if}\,\,\,\, h_{j+1}>x\geq h_j \\
0\,\,\,\,\,\,\,\,\,\,\,\,\,\,\,\,\,\,\,\,\,\,\,\,\,\,\,\,\,\,\textrm{otherwise}.\\
\end{array}
\end{equation}
Eq. (\ref{phiLinear}) can be rearranged as follows:
\begin{equation}
\phi'(H)=\sum_{j=0}^Mu_j\bigg[\frac{h_{j+1}-H}{h_{j+1}-h_j}\Omega_j(H)(1-\delta_{jM})+\frac{H-h_{j-1}}{h_j-h_{j-1}}\Omega_{j-1}(H)(1-\delta_{j0})\bigg].
\end{equation}
Note that the expression in the square brackets is just $\xi_j$ in Eq. (\ref{phiForm}).
Finally, to optimize the parameter set $u_i$, we can define a score function
\begin{equation}\label{D}
D^2=\frac{1}{N}\sum_{i=1}^N\left[R_i^m-R_i^{pl}+\sum_{j=0}^Mu_j\alpha_j({\bf p}_i,R_i^{pl})
-\sum_{j=0}^M\sum_{l=0}^Mu_ju_l\beta_{jl}({\bf p}_i,R_i^{pl})\right]^2,
\end{equation}
where $R_i^{pl}=(R^{pl}*G_{\sigma})({\bf p}_i)$, ${\bf p}_i$ are the $x,y$ coordinates of PB $i$, and $N$ is the number of PBs.

\subsubsection{Optimization and constraints}

There are many ways to chose both the $M$ and the specific values for the $h_i$s. The only guidance
we have in doing so is the trade off between the level of detail in CC$^{opt}$ -- which comes from choosing a large M --
and the robustness of our method against errors and noise -- which requires that $M$ be small. With this in mind, we chose $M=5$:
$h_i=\{-120, 16.69, 50, 74, 200\}$, where $h_1$ and $h_5$ mark the boundary of the $HU$ distribution (see Fig. 2 b)), $h_2$ and $h_4$
are the values at which the $HU$ distribution is $50\%$ of its maximum, which occurs at $h_3$.
In order to avoid unrealistic solutions, 
we need to impose constraints on the way CC$^{opt}$ can behave.
Based on experimental evidence, CC$^{v}$ should not differ from CC$^{pl}$ by more than 5$\%$ \cite{Yang2} for any given HU. 
Hence our first constraint.
The second constraint is on the derivative of the CC.
Some studies suggest that the slope is always positive \cite{Schneider, Doolan, Ainsley}; however, other studies report that the CC can in fact
have the opposite trend locally \cite{Hudobivnik}. In addition, the CC can also be multi-valued. 
With the apparent complexity of a CC in mind, we first apply the constraint that the slope $[$CC$(h_{i+1})-$CC$(h_i)]/(h_{i+1}-h_i)=0$, 
and then relax it by allowing the solpe to be $>-0.0005n$ for $n=1,...,10$. We take as the final CC the average of the
eleven CCs yielded by this procedure.

\subsubsection{Set-up errors and noise}

The quality of the range measurements is subject to various errors and sources of noise. Chief among them are 1) the set-up error, 
which can be as large as $\pm$3mm \cite{Letourneau}; 
2) the set-up of the range measuring device, which we assume to have uncertainty of 2 mm; and 
3) the noise in the range data measured by the device, which, e. g. for a Prompt gamma can be of
the order of 2 mm \cite{Janssens}. To stress our method, we chose it to be 3 mm.
  
In order to test the robustness of our method against patient set-up errors, measurement
noise, and measurement setup errors, we computed the range from
Eq. (\ref{WEPL}) using CC$^{v}$, performed a convolution around the center of each PB with the center shifted by a vector ${\bf s}=(s_x,s_y)$,
$(R^v*G_{\sigma})(x+s_x,y+s_y)$, and  
and added a random variable $\eta$,
sampled from a Normal distribution $\mathcal{N}(r,\eta)$, where $\eta=3$ mm, and
$r$ is a random variable sampled from another Normal distribution $\mathcal{N}(0,2)$.
Hence, $\eta$ represents the random uncertainty of the measurement for each PB, and
$r$ is the random set-up error of the range-measuring device.
The variables $s_x$ and $s_y$ are the set-up errors in the x and y-direction respectively; they were sampled from a
Normal distribution $\mathcal{N}(\mu,\sigma)$ with $\sigma=3$ mm and $\mu$ that was sampled from another Normal
distribution $\mathcal{N}(0,\sigma)$, representing a systematic error.

Because of the errors present in the setup and in the range measurement, optimizing the score
function (\ref{D}) for each fraction, using the range data from that fraction alone, may not lead to
a reliable CC$^{opt}$. It is even conceivable that CC$^{opt}$ may fit CC$^{v}$ worse than CC$^{pl}$ does. 
For this reason, it may be necessary to accumulate range
data over several fractions, thus averaging out the noise and, to some degree, the set-up errors.
Hence, the measured range, $R^m_i$ in the score function must read 
\begin{equation}\label{NoiseFilter}
R^m_i=\frac{1}{K}\sum_{k=1}^KR^m_i(k),
\end{equation} 
where $R^m_i(k)$ is the measured range of PB $i$ for fraction $k$, and $K$ is the number of fractions already delivered.

\subsubsection{Method of optimization}

The search for the optimum set $(u_1,...,u_{5})$ was performed on Mathematica using the ``Nminimize"
function with the ``DifferentialEvolution" method. The aforementioned constraints were stated explicitly in the
``Nminimize" function.

\section{Results}

\begin{figure}
\centering
\includegraphics[trim=0 0 0 0.5cm, height=0.9\textheight]{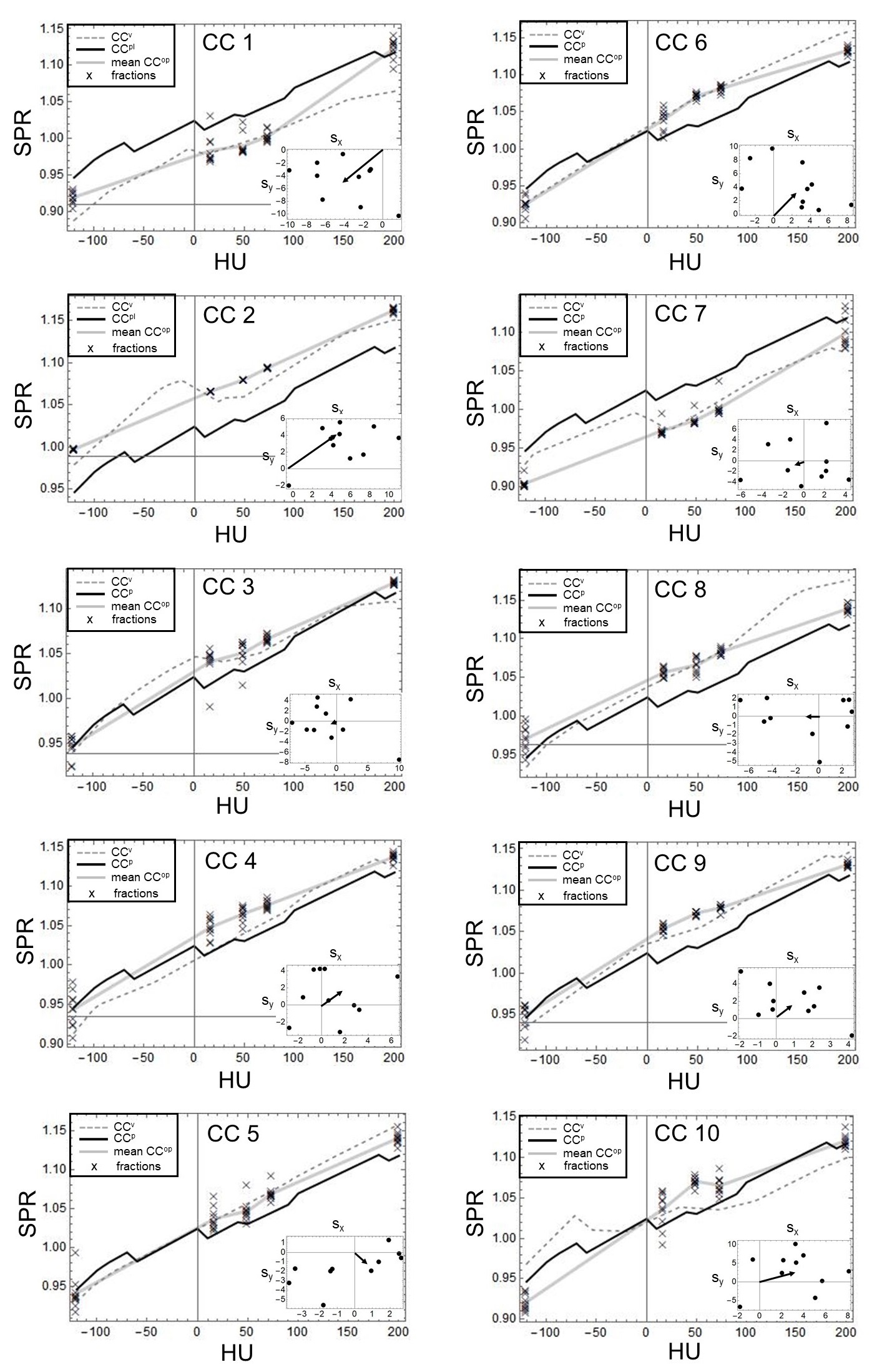}
\caption{CC$^v$ (dashed gray), CC$^{pl}$ (black solid) and CC$^{op}$ for the first ten fractions (crosses). The frames in the lower right corner
show the systematic (arrow) and random (black dot) setup errors for the first ten fractions.}
\end{figure}

In Fig. 3, we show the ten CC$^v$s from Fig. 2a) (dashed), CC$^{pl}$ (solid black) and the CC$^{opt}$ for ten fractions (marked by ``x").
The gray line connects the average values over ten fractions for each of the five HUs. The smaller frames in the bottom left corner
show the x- and y-components of the set-up error in millimeters for ten fractions; the systematic set-up error is indicated by the black arrow. 
By visual inspection, we see that not all
CC$^{opt}$ are an improvement on the CC$^{pl}$. For instance, in CC 4 the average CC$^{opt}$ is too large in the
domain $[0,100]$, where it really counts. In CC 10 we see a similar behavior; however, this case is even worse: in the domain $[-120,0]$ the average
CC$^v$ is below CC$^{pl}$ when it should be above.

Let us go beyond mere visual inspection and look at the average relative deviation of CC$^{opt}$ and CC$^{pl}$ from CC$^v$, defined
by
\begin{equation}\label{SPRunc}
\Delta SPR=\frac{1}{320}\int_{-120}^{200}dy\left[1-\frac{CC^{pl/opt}(y)}{CC^{v}(y)}\right].
\end{equation}
Fig. 4a) shows this quantity for CC$^{pl}$ (black dots) and for CC$^v$ for ten fractions (marked by ``x");
the center of the circle indicates the 5$^{\textrm{th}}$ fraction, while the tip of the triangle overlaps with the 10$^{\textrm{th}}$ fraction.
This figure reveals that, for the 5$^{\textrm{th}}$ fraction, CC$^{opt}$ in CC 4 is indeed worse on average than CC$^{pl}$. However,
the converse is true for the 10$^{\textrm{th}}$ fraction; CC$^{opt}$ gives about 50$\%$ improvement over CC$^{pl}$. By the measure
defined in Eq. (\ref{SPRunc}), CC$^{opt}$ in CC 10, although the best among the ten fractions, is still worse
than CC$^{pl}$. 

The measure that is of real interest is the range; more specifically the distribution of the over/under-range defined as $\sum_{x,y}[R^{v}(x,y)-R^{pl/opt}(x,y)]$.
Fig. 4b) shows the average over/under-range defined as
\begin{equation}\label{AvDelR}
\langle\Delta R^{pl/opt}\rangle=\frac{1}{L_EL_p}\sum_{m=1}^{L_E}\sum_{x,y}\left[R^v(x,y)-R^{pl/opt}(x,y)\right],
\end{equation}
where $L_p$ is the total number of pixels in the sum and $L_E=20$ the number of layers. The area covered by the sum was chosen so as to fit all PBs of the plan, plus
ten additional millimeters surrounding the boarder in anticipation of patient set-up errors. 
In some anatomical regions, e.g. near the interface between the skin and the air, even a small difference
between the CC$^v$ and CC$^{opt}$ can result in either $R^v$ or $R^{pl/opt}$ to traverse the tissue, giving a range equal to the limit of the CT grid. 
Computing $R^v(x,y)-R^{pl/opt}(x,y)$ for such pixels would not be a fair assessment of the effect coming from the difference between CC$^v$ and
CC$^{pl/opt}$. For this reason they were omitted from the sum.
Finally, Fig. 4c) shows the standard deviation of the over/under-range defined as
\begin{equation}\label{DevDelR}
\langle\Delta\Delta R^{pl/opt}\rangle=\left[\frac{1}{L_EL_p}\sum_{m=1}^{L_E}\sum_{x,y}\left[R^v(x,y)-R^{pl/opt}(x,y)-\langle\Delta R^{pl/opt}\rangle\right]^2\right]^{1/2},
\end{equation}
The results of Figs. 4 a), b) and c) are shown explicitly in Table 1.

\begin{figure}
\centering
\includegraphics[trim=0 0 0 0.5cm, height=0.45\textheight]{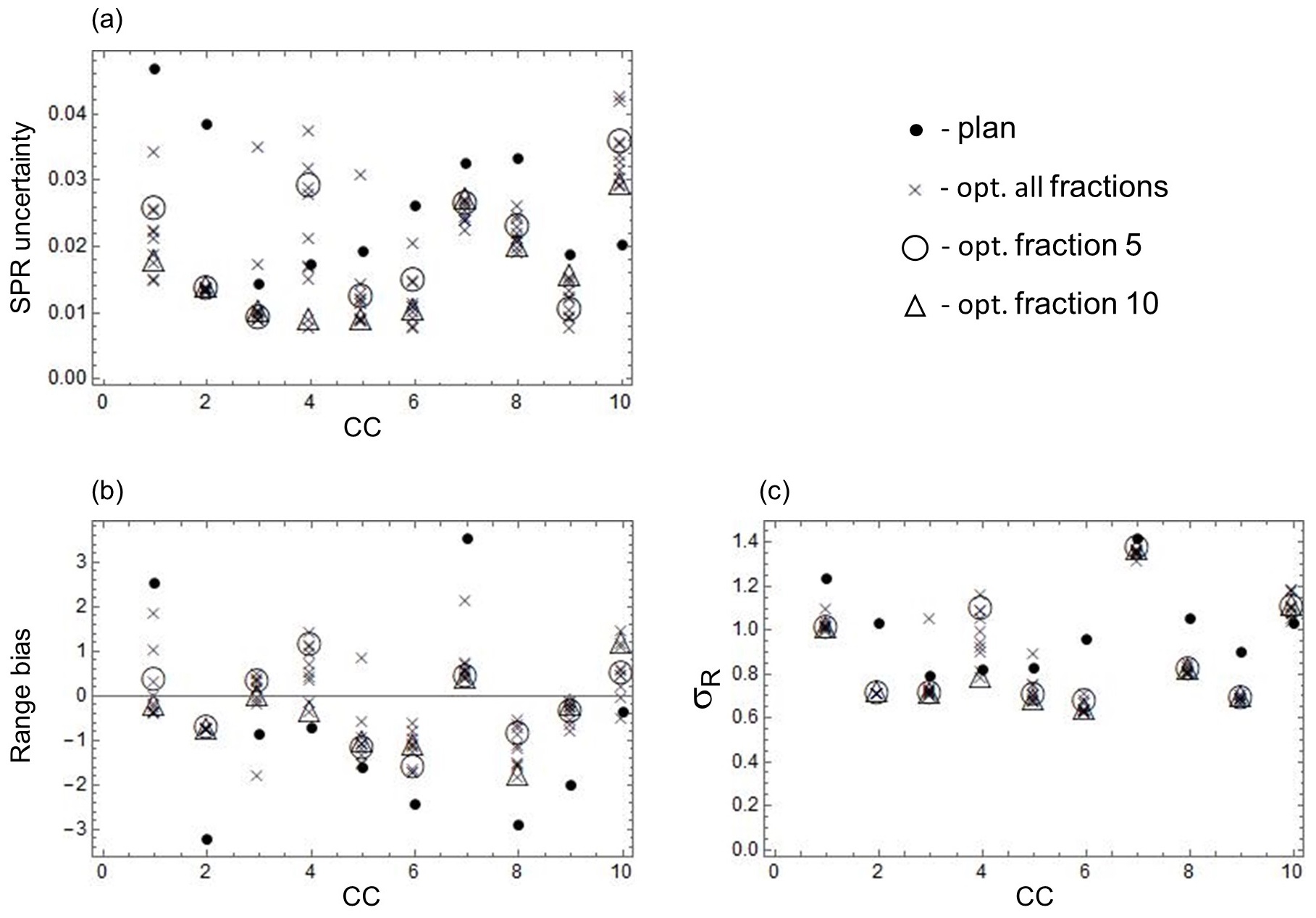}
\caption{a) RSP uncertainty as defined in Eq. (\ref{SPRunc}). b) Range bias and c) range uncertainty as defined in Eqs. (\ref{AvDelR}) and (\ref{DevDelR}), respectively.}
\end{figure}

\begin{table}
 \centering
  \caption{This table shows explicitly the values from Figure 4 a) b) and c).}
  \begin{tabularx}{\textwidth}{XXXXXXXX}
    \hline\hline
    		\footnotesize CC		  & \footnotesize CC$^{pl}$ error & \footnotesize CC$^{opt}$ error & \footnotesize $R^{pl}$ bias & \footnotesize $R^{opt}$ bias & \footnotesize $\sigma_R^{pl}$ & \footnotesize $\sigma_R^{opt}$ \\
				  &  & \footnotesize (frac. 5,\,10) &  & \footnotesize (frac. 5,\,10) & & \footnotesize (frac. 5,\,10) \\
    \hline\hline
    \footnotesize 1 & \footnotesize 4.68\,\% & \footnotesize 2.57,\,\,1.77\,\% & \footnotesize 2.52 & \footnotesize 0.35,\,\,-0.23 & \footnotesize 1.24 & \footnotesize 1.01,\,\,1.01 \\
    \footnotesize 2 & \footnotesize 3.85\,\% & \footnotesize 1.37,\,\,1.38\,\% & \footnotesize -3.24 & \footnotesize  -0.73,\,\,-0.75 & \footnotesize 1.03 & \footnotesize 0.71,\,\,0.71 \\
    \footnotesize 3 & \footnotesize 1.44\,\% & \footnotesize 0.92,\,\,1.02\,\% & \footnotesize -0.87 & \footnotesize 0.32,\,\,-0.02 & \footnotesize 0.80 & \footnotesize 0.71,\,\,0.71 \\
    \footnotesize 4 &  \footnotesize 1.73\,\% &  \footnotesize 2.90,\,\,0.90\,\% & \footnotesize  -0.71 & \footnotesize 1.12,\,\,-0.35 & \footnotesize 0.82 & \footnotesize 1.09,\,\,0.79 \\
    \footnotesize 5 & \footnotesize 1.93\,\% & \footnotesize 1.23,\,\,1.89\,\% & \footnotesize  -1.63 & \footnotesize -1.18,\,\,-1.05 & \footnotesize 0.82 & \footnotesize 0.70,\,\,0.68 \\
    \footnotesize 6 & \footnotesize 2.61\,\% & \footnotesize 1.48,\,\,1.03\,\% & \footnotesize -2.45 & \footnotesize -1.62,\,\,-1.12 & \footnotesize 0.96 & \footnotesize 0.67,\,\,0.64 \\
    \footnotesize 7 & \footnotesize 3.24\,\% & \footnotesize 2.63,\,\,2.72\,\% & \footnotesize 3.53 & \footnotesize 0.40,\,\,0.40 & \footnotesize 1.42 & \footnotesize 1.34,\,\,1.37 \\
    \footnotesize 8 & \footnotesize 3.32\,\% & \footnotesize 2.30,\,\,1.99\,\% & \footnotesize -2.90 & \footnotesize -0.86,\,\,-1.81 & \footnotesize 1.05 & \footnotesize 0.82,\,\,0.82 \\
    \footnotesize 9 & \footnotesize 1.88\,\% & \footnotesize 1.03,\,\,1.55\,\% & \footnotesize -2.00 & \footnotesize -0.35,\,\,-0.21 & \footnotesize 0.90 & \footnotesize 0.68,\,\,0.69 \\
    \footnotesize 10 & \footnotesize 2.02\,\% &  \footnotesize 3.57,\,\,2.93\,\% & \footnotesize  -0.36 & \footnotesize 0.48,\,\,1.19 & \footnotesize  1.03 & \footnotesize 1.11,\,\,1.11 \\
    \hline\hline
  \end{tabularx}
\end{table}

\section{Discussion}

The method of optimizing a CC presented in this paper was tested against various sources of error and noise.
Patient set-up errors, systematic and random, in the plane transverse to the beam, both with $\sigma=3$ mm, were added.
The simulated virtual range of each PB, representing the range as measured by a range-measuring device, was skewed by a
a value sampled from a normal distribution with $\sigma=3$ mm; this value was meant to simulate the uncertainty
in the range measurement. Another source of error with $\sigma=2$ mm was added to capture the
error in the set-up of the measuring device.

Due to so much uncertainty, one cannot expect this method to be reliable for
a given fraction but rather must be applied after several fractions, taking in as input range data that have been
averaged and hence partly filtered. This filtering, however, only applies to random errors, not systematic ones. Nevertheless,
according to Figures 4 a), b) and c), even in the presence of systematic errors the CC, the range bias and the range uncertainty
tend to improve after ten fractions. Even after five fractions, in eight out of ten cases, the average deviation of CC$^{opt}$
from CC$^{v}$ was lower than the average deviation of CC$^{pl}$
from CC$^{v}$. Same was true for the range bias and the range uncertainty. Regarding the range bias in the last case, CC 10,
the fifth fraction is actually better than the tenth fraction. In addition, for several fractions (marked by the crosses), the range bias is close to zero.
This suggests that for this particular case of a CC$^v$, our method is less robust to noise and errors.

The success of our method is on a par with other published methods of optimizing the CC. For example,
it has been demonstrated [ref] that the use of dual energy CT (DECT) can reduce uncertainty in the
stopping power from 1.59$\%$ to 0.61$\%$, and offer a reduction in bias from -0.88$\%$ to -0.58$\%$ and -0.14$\%$.
On average, our method reduced the error in a CC from 2.67$\%$ to 1.62$\%$ and the WEPL bias from 2.14$\%$ to
0.74$\%$.
Other studies... plus a discussion of advantages/disadvantages, e.g. no additional dose to patient.

\section{Conclusions}

The method of optimizing the HU-SP calibration curve presented herein can provide a new way
for clinicians to monitor and/or adapt the course of a treatment.
The main novelty of this method lies in its structure: the bulk of the computation is performed before the treatment so
that the computation required upon the range data acquisition is very efficient (a few seconds).
What makes this possible is the fact that our method relies on an analytical expression, rather than an algorithm,
which takes as input the planning CT, a planning (usually a stoichiometric) CC, and the range data, and yields the optimized CC as output.
Controlling the quality of this output are constraints placed upon the minimization procedure.
We have shown that our method, when applied over several fractions, yields a CC that better serves the patient, i. e. 
a CC that results in the overall reduction of the over/under-range. Although more work is required to ascertain the true
potential of this method, we submit that the present study sufficiently demonstrates its usefulness in
proton radiotherapy.

Lastly, it is worth reiterating that our method is based on an analytical formula whose only inputs are a planning CT
and range data. As such, it is in principle able to incorporate any range measuring device/method, e. g. proton radiography, Prompt gamma camera, PET scan,
ionoacoustic range measurements. The quality of results yielded by our method will, of course, depend on the quality of the range data provided by the
range measuring device. The second advantage of having an analytical formula as a basis is the speed with which
the CC can be optimized. While direct methods, such as Monte Carlo, are preferable to others due to their accuracy, they are
not practical in cases where optimization of many parameters is desired. This is where our method can be of great benefit.

%
%

\section{Acknowledgments}
This research project was made possible by ImagX-R research program co-funded IBA.
JA would like to thank E Ejkova for her technical support.

\end{document}